\documentclass[prd,aps,floatfix,nofootinbib,11 pt]{revtex4}
\usepackage{amsmath,graphicx,color,epsfig}

\input{tcilatex}

\begin{document}

\title{Noise effects in a three-player Prisoner's Dilemma quantum game}
\author{M. Ramzan\thanks{%
mramzan@phys.qau.edu.pk} and M. K. Khan}
\address{Department of Physics Quaid-i-Azam University \\
Islamabad 45320, Pakistan}

\begin{abstract}
We study the three-player Prisoner's Dilemma game under the effect of
decoherence and correlated noise. It is seen that the quantum player is
always better off over the classical players. It is also seen that the
game's Nash equilibrium does not change in the presence of correlated noise
in contradiction to the effect of decoherence in multiplayer case.
Furthermore, it is shown that for maximum correlation the game does not
behave as a noiseless game and the quantum player is still better off for
all values of the decoherence parameter $p$ which is not possible in the
two-player case. In addition, the payoffs reduction due to decoherence is
controlled by the correlated noise throughout the course of the game.\newline
\end{abstract}

\maketitle

\address{Department of Physics Quaid-i-Azam University \\
Islamabad 45320, Pakistan}

\address{Department of Physics Quaid-i-Azam University \\
Islamabad 45320, Pakistan}


\vspace*{1.0cm}

\vspace*{1.0cm}



\section{Introduction}

Quantum entanglement provides a fundamental potential resource for
communication and information processing and is one of the key quantitative
notions of the intriguing field of quantum information theory and quantum
computation. A quantum superposition state decays into a classical,
statistical mixture of states through a decoherence process which is caused
by entangling interactions between the system and its environment \cite{c}.
Superposition of quantum states however, are very fragile and easily
destroyed by the decoherence processes. Such uncontrollable influences cause
noise in the communication or errors in the outcome of a computation, and
thus reduce the advantages of quantum information methods. However, in a
more realistic and practical situation, decoherence caused by an external
environment is inevitable. Therefore, influence of an external environmental
system on the entanglement cannot be ignored. A novel research has been
carried out to study the quantum communication channels. Macchiavello and
Palma \cite{bb} have developed the theory of quantum channels to encompass
memory effects. In real-world applications the assumption of having
uncorrelated noise channels can not be fully justified. However, quantum
computing in the presence of noise is possible with the use of decoherence
free subspaces \cite{k} and the quantum error correction \cite{Pres}.

Application of mathematical physics to economics has seen a recent
development in the form of quantum game theory. Two-player quantum games
have attracted a lot of interest in recent years [5-7]. A number of authors
have investigated the quantum prisoner's dilemma game [8-10]. A detailed
description on quantum game theory can be found in references [11-16]. There
have been remarkable advances in the experimental realization of quantum
games such as Prisoner's Dilemma \cite{zhu,a}. The Prisoner's Dilemma game
is a widely known example in classical game theory. The quantum version of
the Prisoner's Dilemma has been experimentally demonstrated using a nuclear
magnetic resonance (NMR) quantum computer \cite{a}. Recently, Prevedel et
al. have experimentally demonstrated the application of a measurement-based
protocol \cite{b}. They realized a quantum version of the Prisoner's Dilemma
game based on the entangled photonic cluster states. It was the first
realization of a quantum game in the context of one-way quantum computing.
Studies concerning the quantum games in the presence of decoherence and
correlated noise have produced interesting results. Chen et al. \cite{I}\
have shown that in the case of two-player Prisoner's Dilemma game, the Nash
equilibria are not changed by the effect of decoherence in a maximally
entangled case. Nawaz and Toor \cite{m} have studied quantum games under the
effect of correlated noise by taking a particular example of the
phase-damping channel. They have shown that the quantum player outperforms
the classical players for all values of the decoherence parameter $p$. They
have also shown that for maximum correlation the effects of decoherence
diminish and it behaves as a noiseless game. Recently, we have investigated
different quantum games under different noise models and found interesting
results \cite{n}. More recently, Gawron et al. \cite{QMSG} have studied the
noise effects in quantum magic squares game. They have shown that the
probability of success can be used to determine characteristics of quantum
channels. Investigation of multiplayer quantum games in a multi-qubit system
could be of much interest and significance. In the recent years, quantum
games with more than two players were investigated [24-27]. Such games can
exhibit certain forms of pure quantum equilibrium that have no analog in
classical games, or even in two-player quantum games. Recently, Cao et al. 
\cite{rr} have investigated the effect of quantum noise on a multiplayer
Prisoner's Dilemma quantum game. They have shown that in a maximally
entangled case a special Nash equilibrium appears for a specific range of
the quantum noise parameter (the decoherence parameter). However, yet no
attention has been given to the multiplayer quantum games under the effect
of correlated noise, which is the main focus of this paper.

In this paper, we investigate three-player Prisoner's Dilemma quantum game
under the effect of decoherence and correlated noise in a three-qubit
system. We have considered a dephasing channel parameterized by the memory
factor $\mu $ which measures the degree of correlations. By exploiting the
initial state and measurement basis entanglement parameters, $\gamma \in
\lbrack 0,\pi /2]$ and $\delta \in \lbrack 0,\pi /2],$ we study the role of
decoherence parameter $p\in \lbrack 0,1]$ and memory parameter $\mu \in
\lbrack 0,1]$ on the three-player Prisoner's Dilemma quantum game. Here, $%
\delta =0$ means that the measurement basis are unentangled and $\delta =\pi
/2$ means that it is maximally entangled. Similarly, $\gamma =0$ means that
the game is initially unentangled and $\gamma =\pi /2$ means that it is
maximally entangled. Whereas the lower and upper limits of $p$ correspond to
a fully coherent and fully decohered system, respectively. Similarly, the
lower and upper limits of $\mu $ correspond to a memoryless and maximum
memory (degree of correlation) cases, respectively. It is seen that in
contradiction to the two-player Prisoner's Dilemma quantum game, in the
three-player game, the quantum player can outperform the classical players
for all values of the decoherence parameter $p$ for the maximum degree of
correlations (i.e. memory parameter $\mu $ $=1$). In comparison to the
two-player situation, the three-player game does not become noiseless and
quantum player still remains superior over the classical ones for an entire
range of the decoherence parameter, $p,$ in memoryless case i.e. $\mu =0$.
It is shown that the payoffs reduction due to decoherence is controlled by
the memory parameter $\mu $ throughout the course of the game. It is also
shown that the Nash equilibrium of the game does not change under the
correlated noise in contradiction to the case of decoherence effects as
investigated by Cao et al. \cite{rr}.

\section{Three-player Prisoner's Dilemma game}

Properties of the two-player quantum games have been discussed extensively
[11-13, 29], however, not much attention has been given to the multiplayer
quantum games. Study of the multiplayer games may exhibit interesting
results in comparison to the two-player games. The three-player Prisoners'
Dilemma is similar to the two-player situation except that Alice, Bob and a
third player Charlie join the game. The three players are arrested under the
suspicion of robbing a bank. Similar to two-player case, they are
interrogated in separate cells without communicating with each other. The
two possible moves for each prisoner are, to cooperate $(C)$ or to defect $%
(D).$ The payoff table for the three-player Prisoner's Dilemma is shown in
table 1. The game is symmetric for the three players, and the strategy $D$
dominates the strategy $C$ for all of them. Since the selfish players prefer
to choose $D$ as the optimal strategy, the unique Nash equilibrium is ($%
D,D,D $) with payoffs ($1,1,1$). This is a Pareto inferior outcome, since ($%
C,C,C$) with payoffs ($3,3,3$) would be better for all the three players.
This situation is the very catch of the dilemma and is similar to the
two-player version of this game. The dilemma of this game can be resolved in
its quantum version. Du et al. [25] investigated the three-player quantum
Prisoner's Dilemma game with a certain strategic space. They found a Nash
equilibrium that can remove the dilemma in the classical game when the
game's state is maximally entangled. This particular Nash equilibrium
remains to be a Nash equilibrium even for the non-maximally entangled cases.
However, their calculations for the expected payoffs of the players comprise
product measurement basis for the arbiter of the game. Here in our model we
use the entangled measurement basis for the arbiter of the game to perform
measurement. In addition, we include the effect of decoherence and
correlated noise in the three-players settings.

\section{Time correlated dephasing channel}

Quantum information is encoded in qubits during its transmission from one
party to another and requires a communication channel. In a realistic
situation, the qubits have a nontrivial dynamics during transmission because
of their interaction with the environment. Therefore, Bob may receive a set
of distorted qubits because of the disturbing action of the channel. Studies
on quantum channels have attracted a lot of attention in the recent years 
\cite{bb,oo}. Early work in this direction was devoted mainly, to memoryless
channels for which consecutive signal transmissions through the channel are
not correlated. In the correlated channels (channels with the memory), the
noise acts on consecutive uses of channels. We consider here the noise model
based on the time correlated dephasing channel. In the operator sum
representation, the dephasing process can be expressed as \cite{p} 
\begin{equation}
\rho _{f}=\sum\limits_{i=0}^{1}A_{i}\rho _{in}A_{i}^{\dagger }
\end{equation}%
where 
\begin{eqnarray}
A_{0} &=&\sqrt{1-\frac{p}{2}}I  \notag \\
A_{1} &=&\sqrt{\frac{p}{2}}\sigma _{z}
\end{eqnarray}%
are the Kraus operators, $I$ is the identity operator $\sigma _{z}$ is the
Pauli matrix and $p$ is the decoherence parameter. Let $N$ qubits are
allowed to pass through such a channel then equation (1) becomes \cite%
{Flitney2}%
\begin{equation}
\rho _{f}=\sum\limits_{k_{1,}....,.k_{n}=0}^{1}(A_{k_{n}}\otimes
.....A_{k_{1}})\rho _{in}(A_{k_{1}}^{\dagger }\otimes
.....A_{k_{n}}^{\dagger })
\end{equation}%
Now if the noise is correlated with the memory of degree $\mu ,$ then the
action of the channel on the two consecutive qubits is given by the Kraus
operators \cite{bb}%
\begin{equation}
A_{ij}=\sqrt{p_{i}[(1-\mu )p_{j}+\mu \delta _{ij}]}\sigma _{i}\otimes \sigma
_{j}
\end{equation}%
where $\sigma _{i}$ and $\sigma _{j}$ are usual Pauli matrices with indices $%
i$ and $j$ run from $0$ to $3$ and $\mu $ is the memory parameter$.$ The
above expression means that with the probability $(1-%
\mu
)$ the noise is uncorrelated whereas with the probability $%
\mu
$ the noise is correlated. Physically the parameter $%
\mu
$ is determined by the relaxation time of the channel when a qubit passes
through it. In order to remove correlations, one can wait until the channel
has relaxed to its original state before sending the next qubit. However,
this may lower the rate of information transfer. The Kraus operators for the
three qubit system can be written as \cite{qqq}%
\begin{equation}
A_{ijk}=\sqrt{[(1-\mu )p_{i}+\mu \delta _{ij}][(1-\mu )p_{j}+\mu \delta
_{jk}]p_{k}}\sigma ^{i}\otimes \sigma ^{j}\otimes \sigma ^{k}
\end{equation}%
where $i,$ $j,$ $k$ are $0$ or $3.$ The memory parameter $%
{\mu}%
$ is contained in the probabilities $A_{ijk},$ which determines the
probability of the errors $\sigma ^{i}\otimes \sigma ^{j}\otimes \sigma
^{k}. $ Recalling that $(1-%
{\mu}%
)$ is the probability of independent errors on two consecutive qubits and $%
{\mu}%
$ is the probability of identical errors. The sum of probabilities of all
types of errors on the three qubits add to unity as we expect,%
\begin{equation}
\sum\limits_{i,j,k}[(1-\mu )^{2}A_{i}A_{j}A_{k}+2\mu (1-\mu )A_{i}A_{j}+\mu
^{2}A_{i}]=1
\end{equation}%
It is necessary to consider the performance of the channel for arbitrary
values of the $%
\mu
$ to reach a compromise between various factors which determine the final
rate of information transfer.\ Thus in passing through the channel any two
consecutive qubits undergo random independent (uncorrelated) errors with the
probability ($1-%
\mu
)$ and identical (correlated) errors with the probability $%
\mu
$. This should be the case if the channel has a memory depending on its
relaxation time and if we stream the qubits through it.

\section{The model}

In our model, Alice, Bob and Charlie, each uses individual channels to
communicate with the arbiter of the game. The two uses of the channel i.e.
the first passage (from the arbiter) and the second passage (back to the
arbiter) are correlated as depicted in figure 1. We consider that the
initial entangled state is prepared by the arbiter and passed on to the
players through a quantum correlated dephasing channel (QCDC). On receiving
the quantum state, the players apply their local operators (strategies) and
return it back to the arbiter via QCDC. Then, the arbiter performs the
measurement and announces their payoffs. Let's consider that the three
players Alice, Bob and Charlie be given the following initial quantum state:%
\begin{equation}
\left\vert \psi _{in}\right\rangle =\cos \frac{\gamma }{2}\left\vert
000\right\rangle +i\sin \frac{\gamma }{2}\left\vert 111\right\rangle
\end{equation}%
where $0\leq \gamma \leq \pi /2$ corresponds to the entanglement of the
initial state. The players can locally manipulate their individual qubits.
The strategies of the players can be represented by the unitary operator $%
U_{i}$ of the form \cite{n}. 
\begin{equation}
U_{i}=\cos \frac{\theta _{i}}{2}R_{i}+\sin \frac{\theta _{i}}{2}P_{i}
\end{equation}%
where $i=1,$ $2$ or $3$\ and $R_{i}$, $P_{i}$ are the unitary operators
defined as 
\begin{eqnarray}
R_{i}\left\vert 0\right\rangle &=&e^{i\alpha _{i}}\left\vert 0\right\rangle
,\qquad \qquad R_{i}\left\vert 1\right\rangle =e^{-i\alpha _{i}}\left\vert
1\right\rangle  \notag \\
P_{i}\left\vert 0\right\rangle &=&e^{i(\frac{\pi }{2}-\beta _{i})}\left\vert
1\right\rangle ,\qquad P_{i}\left\vert 1\right\rangle =e^{i(\frac{\pi }{2}%
+\beta _{i})}\left\vert 0\right\rangle
\end{eqnarray}%
where $0\leq \theta _{i}\leq \pi ,$ and $-\pi \leq \{\alpha _{i},$ $\beta
_{i}\}\leq \pi .$ Application of the local operators of the players
transforms the initial state given in equation (7) to 
\begin{equation}
\rho _{f}=(U_{1}\otimes U_{2}\otimes U_{3})\rho _{in}(U_{1}\otimes
U_{2}\otimes U_{3})^{\dagger }
\end{equation}%
where $\rho _{in}=\left\vert \psi _{in}\right\rangle \left\langle \psi
_{in}\right\vert $ is the density matrix for the quantum state. The
operators used by the arbiter to determine the payoffs for Alice, Bob and
Charlie are 
\begin{eqnarray}
P^{k}
&=&\$_{000}^{k}P_{000}+\$_{001}^{k}P_{001}+\$_{110}^{k}P_{110}+%
\$_{010}^{k}P_{010}  \notag \\
&&+\$_{101}^{k}P_{101}+\$_{011}^{k}P_{011}+\$_{100}^{k}P_{100}+%
\$_{111}^{k}P_{111}
\end{eqnarray}%
where $k=A$, $B$ or $C$ and 
\begin{eqnarray}
P_{000} &=&\left\vert \psi _{000}\right\rangle \left\langle \psi
_{000}\right\vert ,\qquad \left\vert \psi _{000}\right\rangle =\cos \frac{%
\delta }{2}\left\vert 000\right\rangle +i\sin \frac{\delta }{2}\left\vert
111\right\rangle  \notag \\
P_{111} &=&\left\vert \psi _{111}\right\rangle \left\langle \psi
_{111}\right\vert ,\qquad \left\vert \psi _{111}\right\rangle =\cos \frac{%
\delta }{2}\left\vert 111\right\rangle +i\sin \frac{\delta }{2}\left\vert
000\right\rangle  \notag \\
P_{001} &=&\left\vert \psi _{001}\right\rangle \left\langle \psi
_{001}\right\vert ,\qquad \left\vert \psi _{001}\right\rangle =\cos \frac{%
\delta }{2}\left\vert 001\right\rangle +i\sin \frac{\delta }{2}\left\vert
110\right\rangle  \notag \\
P_{110} &=&\left\vert \psi _{110}\right\rangle \left\langle \psi
_{110}\right\vert ,\qquad \left\vert \psi _{110}\right\rangle =\cos \frac{%
\delta }{2}\left\vert 110\right\rangle +i\sin \frac{\delta }{2}\left\vert
001\right\rangle  \notag \\
P_{010} &=&\left\vert \psi _{010}\right\rangle \left\langle \psi
_{010}\right\vert ,\qquad \left\vert \psi _{010}\right\rangle =\cos \frac{%
\delta }{2}\left\vert 010\right\rangle -i\sin \frac{\delta }{2}\left\vert
101\right\rangle  \notag \\
P_{101} &=&\left\vert \psi _{101}\right\rangle \left\langle \psi
_{101}\right\vert ,\qquad \left\vert \psi _{101}\right\rangle =\cos \frac{%
\delta }{2}\left\vert 101\right\rangle -i\sin \frac{\delta }{2}\left\vert
010\right\rangle  \notag \\
P_{011} &=&\left\vert \psi _{011}\right\rangle \left\langle \psi
_{011}\right\vert ,\qquad \left\vert \psi _{011}\right\rangle =\cos \frac{%
\delta }{2}\left\vert 011\right\rangle -i\sin \frac{\delta }{2}\left\vert
100\right\rangle  \notag \\
P_{100} &=&\left\vert \psi _{100}\right\rangle \left\langle \psi
_{100}\right\vert ,\qquad \left\vert \psi _{100}\right\rangle =\cos \frac{%
\delta }{2}\left\vert 100\right\rangle -i\sin \frac{\delta }{2}\left\vert
011\right\rangle  \label{mbasis}
\end{eqnarray}%
where $0\leq \delta \leq \pi /2$ and $\$_{lmn}^{k}$ are elements of the
payoff matrix as given in table 1. Since quantum mechanics is a
fundamentally probabilistic theory, the strategic notion of the payoff is
the expected payoff. The players after their actions, forward their qubits
to the arbiter of the game for the final projective measurement in the
computational basis (see equation (\ref{mbasis})). The arbiter of the game
finally determines their payoffs (see figure 1). The payoffs for the players
can be obtained as the mean values of the payoff operators as 
\begin{equation}
\$_{k}(\theta _{i},\alpha _{i},\beta _{i})=\text{Tr}(P^{k}\rho _{f})
\end{equation}%
where Tr represents the trace of the matrix. Using equations (5) to (13),
the payoffs for the three players can be obtained as 
\begin{eqnarray}
&&\left. \$_{k}(\theta _{i},\alpha _{i},\beta _{i})=\right.  \notag \\
&&c_{1}c_{2}c_{3}[\eta _{1}\$_{000}^{k}+\eta
_{2}\$_{111}^{k}+(\$_{000}^{k}-\$_{111}^{k})\mu _{p}^{(1)}\mu _{p}^{(2)}\xi
\cos 2(\alpha _{1}+\alpha _{2}+\alpha _{3})]  \notag \\
&&+s_{1}s_{2}s_{3}[\eta _{2}\$_{000}^{k}+\eta
_{1}\$_{111}^{k}-(\$_{000}^{k}-\$_{111}^{k})\mu _{p}^{(1)}\mu _{p}^{(2)}\xi
\cos 2(\beta _{1}+\beta _{2}+\beta _{3})]  \notag \\
&&+c_{1}c_{2}s_{3}[\eta _{1}\$_{001}^{k}+\eta
_{2}\$_{110}^{k}+(\$_{001}^{k}-\$_{110}^{k})\mu _{p}^{(1)}\mu _{p}^{(2)}\xi
\cos 2(\alpha _{1}+\alpha _{2}-\beta _{3})]  \notag \\
&&+s_{1}s_{2}c_{3}[\eta _{2}\$_{001}^{k}+\eta
_{1}\$_{110}^{k}-(\$_{001}^{k}-\$_{110}^{k})\mu _{p}^{(1)}\mu _{p}^{(2)}\xi
\cos 2(\beta _{1}+\beta _{2}-\alpha _{3})]  \notag \\
&&+s_{1}c_{2}c_{3}[\eta _{1}\$_{100}^{k}+\eta
_{2}\$_{011}^{k}+(\$_{100}^{k}-\$_{011}^{k})\mu _{p}^{(1)}\mu _{p}^{(2)}\xi
\cos 2(\alpha _{2}+\alpha _{3}-\beta _{1})]  \notag \\
&&+c_{1}s_{2}s_{3}[\eta _{2}\$_{100}^{k}+\eta
_{1}\$_{011}^{k}-(\$_{100}^{k}-\$_{011}^{k})\mu _{p}^{(1)}\mu _{p}^{(2)}\xi
\cos 2(\beta _{2}+\beta _{3}-\alpha _{1})]  \notag \\
&&+s_{1}c_{2}s_{3}[\eta _{1}\$_{101}^{k}+\eta
_{2}\$_{010}^{k}+(\$_{101}^{k}-\$_{010}^{k})\mu _{p}^{(1)}\mu _{p}^{(2)}\xi
\cos 2(\beta _{1}+\beta _{3}-\alpha _{2})]  \notag \\
&&+c_{1}s_{2}c_{3}[\eta _{2}\$_{101}^{k}+\eta
_{1}\$_{010}^{k}-(\$_{101}^{k}-\$_{010}^{k})\mu _{p}^{(1)}\mu _{p}^{(2)}\xi
\cos 2(\alpha _{1}+\alpha _{3}-\beta _{2})]  \notag \\
&&+\frac{\mu _{p}^{(1)}}{8}(\cos ^{2}(\delta /2)-\sin ^{2}(\delta
/2))[\$_{000}^{k}-\$_{111}^{k}-\$_{001}^{k}+\$_{110}^{k}-\$_{010}^{k}+%
\$_{101}^{k}+\$_{011}^{k}-\$_{100}^{k}]\times  \notag \\
&&\sin (\gamma )\sin (\theta _{1})\sin (\theta _{2})\sin (\theta _{3})\cos
(\alpha _{1}+\alpha _{2}+\alpha _{3}-\beta _{1}-\beta _{2}-\beta _{3}) 
\notag \\
&&+[[\$_{000}^{k}-\$_{111}^{k}]\sin (\delta )\sin (\theta _{1})\sin (\theta
_{2})\sin (\theta _{2})\cos (\alpha _{1}+\alpha _{2}+\alpha _{3}-\beta
_{1}-\beta _{2}-\beta _{3})  \notag \\
&&+[\$_{110}^{k}-\$_{001}^{k}]\sin (\delta )\sin (\theta _{1})\sin (\theta
_{2})\sin (\theta _{2})\cos (\alpha _{1}+\alpha _{2}-\alpha _{3}+\beta
_{1}+\beta _{2}-\beta _{3})  \notag \\
&&+[\$_{010}^{k}-\$_{101}^{k}]\sin (\delta )\sin (\theta _{1})\sin (\theta
_{2})\sin (\theta _{2})\cos (\alpha _{1}-\alpha _{2}+\alpha _{3}+\beta
_{1}-\beta _{2}+\beta _{3})  \notag \\
&&+[\$_{100}^{k}-\$_{011}^{k}]\sin (\delta )\sin (\theta _{1})\sin (\theta
_{2})\sin (\theta _{2})\cos (\alpha _{1}-\alpha _{2}-\alpha _{3}+\beta
_{1}-\beta _{2}-\beta _{3})]\times  \notag \\
&&[\frac{\mu _{p}^{(2)}}{8}(\cos ^{2}(\gamma /2)-\sin ^{2}(\gamma /2))]
\end{eqnarray}%
where 
\begin{eqnarray}
\mu _{p}^{(j)} &=&(1-p_{j})(1-2p_{j}+4\mu _{j}p_{j}-2\mu
_{j}^{2}p_{j}+p_{j}^{2}-2\mu _{j}p_{j}^{2}+\mu _{j}^{2}p_{j}^{2})  \notag \\
\eta _{1} &=&\cos ^{2}(\gamma /2)\cos ^{2}(\delta /2)+\sin ^{2}(\gamma
/2)\sin ^{2}(\delta /2)  \notag \\
\eta _{2} &=&\{Sin^{2}(\gamma /2)\cos ^{2}(\delta /2)+\sin ^{2}(\delta
/2)\cos ^{2}(\gamma /2)  \notag \\
\ &=&\frac{1}{2}\sin (\delta )\sin (\gamma )\text{ }c_{i}=\cos ^{2}\frac{%
\theta _{i}}{2},s_{i}=\sin ^{2}\frac{\theta _{i}}{2}
\end{eqnarray}%
where $j=1$ or $2$. The payoffs for the three players can be found by
substituting the appropriate values for $\$_{lmn}^{k}$ into equation (14).
Elements of the classical payoff matrix for the Prisoner's Dilemma game are
given in table 1 . The payoff matrix under decoherence can be obtained by
setting $\mu =0$ i.e. by setting $\mu _{p}^{(j)}=(1-p_{j})^{3}$ in equation
(15). It is important to mention that for $p$ and $\mu $\ we mean $%
p_{1}=p_{2}=p$ and $\mu _{1}=\mu _{2}=\mu $\ unless otherwise specified. Our
results are consistent with Ref. [25, 27] and can be verified from equation
(14) when all the three players resort to their Nash equilibrium strategies.
It can be seen that the decoherence causes a reduction in the payoffs of the
players in the memoryless case (see equation (14)). We consider here that
Alice and Bob are restricted to play classical strategies, i.e., $\alpha
_{1}=\alpha _{2}=\beta _{1}=\beta _{2}=0$, whereas Charlie is allowed to
play the quantum strategies as well. It is shown that the quantum player
outperforms the classical players for all values of the decoherence
parameter $p$ for an entire range of the memory parameter $\mu $. Under
these circumstances, it is seen that in contradiction to the two-player
Prisoner's Dilemma\ quantum game, for maximum degree of correlations the
effect of decoherence survives and it does not behave as a noiseless game.
It can be seen that the memory compensates the payoffs reduction due to
decoherence. Further more, it is shown that the memory has no effect on the
Nash equilibrium of the game. Alice's best strategy ($\alpha _{1}=\theta
_{1}=\pi /2,$ and $\beta _{1}=0)$ remains her best strategy throughout the
course of the game. This implies that the correlated noise has no effect on
the Nash equilibrium of the game.

\section{Results and discussions}

To analyze the effects of correlated noise (memory) and decoherence on the
dynamics of the three-player Prisoner's Dilemma quantum game. We consider
the restricted game scenario where Alice and Bob are allowed to play the
classical strategies, i.e., $\alpha _{1}=\alpha _{2}=\beta _{1}=\beta _{2}=0$%
, whereas Charlie is allowed to play the quantum strategies. In figure 2, we
have plotted the players payoffs as a function of the decoherence parameter $%
p$ for the dephasing channel. It is seen that the quantum player out scores
the classical players for all values of the decoherence parameter $p$ for
the memoryless ($\mu =0)$ case. It is shown that even for a maximum degree
of memory i.e. $\mu =1,$ the quantum player can outperform the classical
players, which is in contradiction to the two-player Prisoner's Dilemma\
quantum game. In addition, the decoherence effects persist for maximum
correlation and it does not behave as a noiseless game, contrary to the
two-player case. In figure 3, we have plotted payoffs of the classical and
the quantum players as a function of the memory parameter $\mu $ for $p=0.3$
and $0.7$ respectively.\ It is seen that memory compensates the payoffs
reduction due to decoherence. In figures 4 and 5, we have plotted Alice's
payoff as a function of her strategies $\alpha _{1}$ and $\theta _{1}$ for $%
p=\mu =0.3$ and $p=\mu =0.7$ respectively. It can be seen that the memory
has no effect on the Nash equilibrium of the game. It is evident from
figures 4 and 5 that the best strategy for Alice is $\alpha _{1}=\theta
_{1}=\pi /2,$ and $\beta _{1}=0$. It remains her best strategy for the full
range of the decoherence parameter $p$ and the memory parameter $\mu ,$
throughout the course of the game. Therefore, it can be inferred that
correlated noise has no effect on the Nash equilibrium of the game. In
comparison to the investigations of Cao et al. \cite{rr}, it is shown that
the new Nash equilibrium, appearing for a specific range of the decoherence
parameter $p,$ disappears under the effect of correlated noise. As it can be
seen that for the entire range of the decoherence parameter $p$ and the
memory parameter $\mu ,$ the Nash equilibrium of the game does not change
(see figures 4 and 5). Further more, it can also be seen that the payoffs of
the players are increased with the addition of the correlated noise as can
be seen from figures 4 and 5 respectively, for the entire ranges of the
decoherence and the memory parameters.

\section{Conclusions}

We present a quantization scheme for the three-player Prisoner's Dilemma
game under the effect of decoherence and correlated noise. We study the
effects of decoherence and correlated noise on the game dynamics. We
consider a restricted game situation, where Alice and Bob are restricted to
play the classical strategies, i.e., $\alpha _{1}=\alpha _{2}=\beta
_{1}=\beta _{2}=0$, however Charlie is allowed to play the quantum
strategies as well. It is shown that the quantum player is always better off
for all values of the decoherence parameter $p$ for increasing values of the
memory parameter $\mu $. It is seen that for maximum degree of correlations,
the effect of decoherence does not vanish in comparison to the two-player
Prisoner's Dilemma quantum game. The three-players game doe not become
noiseless game which is in contradiction to the two-player case. It is also
seen that for the maximum degree of memory i.e. $\mu =1$, that the quantum
player can out score the classical players for an entire range of the
decoherence parameter $p$. The payoffs reduction due to the decoherence is
controlled by the memory parameter throughout the course of the game.\
Furthermore, it is shown that the memory has no effect on the Nash
equilibrium of the game.

{\Huge Figures Captions}\newline
\textbf{Figure 1}. Schematic diagram of the model.\newline
\textbf{Figure 2}. Players payoffs as a function of the decoherence
parameter $p$ for the dephasing channel are plotted for the quantum
Prisoner's Dilemma game, with memory parameter $\mu =1$ (solid lines), $\mu
=0$ (dotted lines). $\$_{A(B)}$ are payoffs of the classical players
(Alice/Bob) while $\$_{C}$ represents the payoff of the quantum player
(Charlie). The other parameters are $\theta _{1}=\theta _{2}=\theta _{3}=\pi
/2,$ $\beta _{1}=\beta _{2}=\alpha _{1}=\alpha _{2}=0,$ $\delta =\gamma =\pi
/2$ and $\alpha _{3}=\pi /2,$ $\beta _{3}=\pi /2$ are the optimal strategies
of Charlie.\newline
\textbf{Figure 3}. Payoffs of the classical players (Alice /Bob) and the
quantum player (Charlie) are plotted as a function of memory parameter $\mu
. $ $a_{1}$ and $a_{2}$ are payoffs of the classical players for values of
the decoherence parameter $p=0.7$ and $p=0.3$ respectively. $c_{1}$ and $%
c_{2}$ are payoffs of quantum player for $p=0.7$ and $p=0.3$ respectively.
The other parameters are $\theta _{1}=\theta _{2}=\theta _{3}=\pi /2,$ $%
\beta _{1}=\beta _{2}=\alpha _{1}=\alpha _{2}=0,$ $\delta =\gamma =\pi /2$
and $\alpha _{3}=\pi /2,$ $\beta _{3}=\pi /2$ are the optimal strategies of
Charlie.\newline
\textbf{Figure 4}. Alice's payoff is plotted as a function of her strategies 
$\alpha _{1}$ and $\theta _{1}$ with $\theta _{2}=\theta _{3}=\pi /2,$ $%
\alpha _{2}=\alpha _{3}=\beta _{1}=\beta _{2}=\beta _{3}=0,$ $\delta =\gamma
=\pi /2$ and $p=\mu =0.3.$\newline
\textbf{Figure 5}. Alice's payoff is plotted as a function of her strategies 
$\alpha _{1}$ and $\theta _{1}$ with $\theta _{2}=\theta _{3}=\pi /2,$ $%
\alpha _{2}=\alpha _{3}=\beta _{1}=\beta _{2}=\beta _{3}=0,$ $\delta =\gamma
=\pi /2$ and $p=\mu =0.7.$\newline

{\Huge Table Caption}\newline
\textbf{Table 1}. The payoff matrix for the three-player Prisoner's Dilemma
game where the first number in the parenthesis denotes the payoff of Alice,
the second number denotes the payoff of Bob and the third number denotes the
payoff of Charlie.\newpage

\begin{figure}[tbp]
\begin{center}
\vspace{-2cm} \includegraphics[scale=0.6]{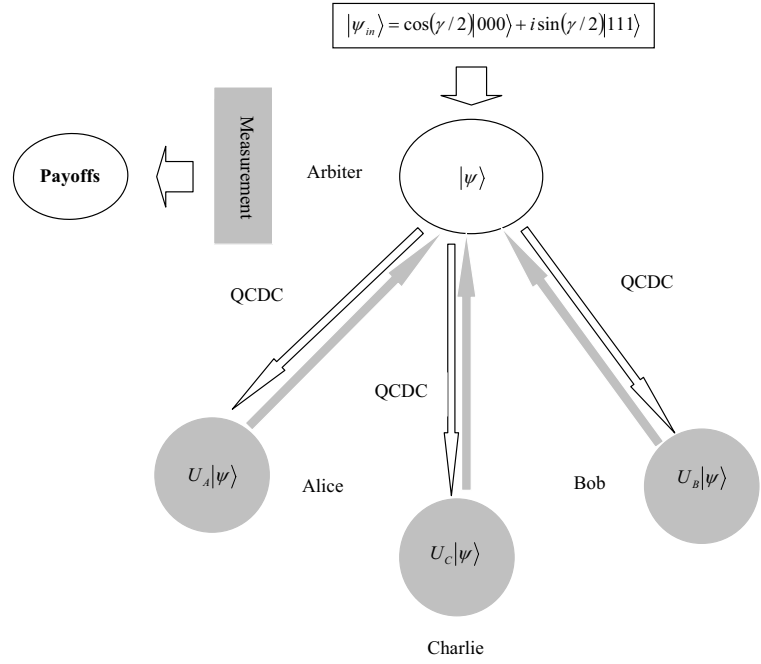} \\[0pt]
\end{center}
\caption{Schematic diagram of the model.}
\end{figure}

\begin{figure}[tbp]
\begin{center}
\vspace{-2cm} \includegraphics[scale=0.6]{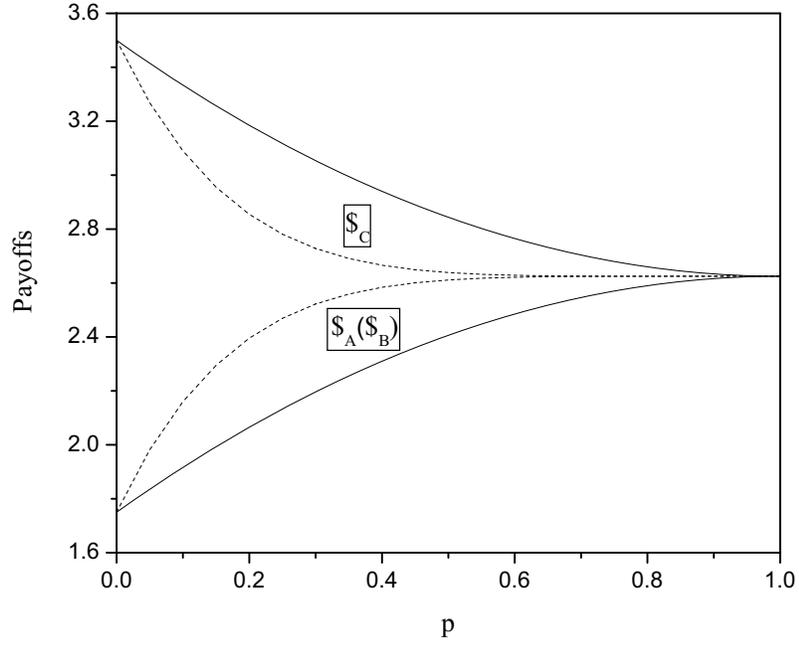} \\[0pt]
\end{center}
\caption{Players payoffs as a function of the decoherence parameter $p$ for
the dephasing channel are plotted for the quantum Prisoner's Dilemma game,
with memory parameter $\protect\mu =1$ (solid lines), $\protect\mu =0$
(dotted lines). $\$_{A(B)}$ are payoffs of the classical players (Alice/Bob)
while $\$_{C}$ represents the payoff of the quantum player (Charlie). The
other parameters are $\protect\theta _{1}=\protect\theta _{2}=\protect\theta %
_{3}=\protect\pi /2,$ $\protect\beta _{1}=\protect\beta _{2}=\protect\alpha %
_{1}=\protect\alpha _{2}=0,$ $\protect\delta =\protect\gamma =\protect\pi /2$
and $\protect\alpha _{3}=\protect\pi /2,$ $\protect\beta _{3}=\protect\pi /2$
are the optimal strategies of Charlie.}
\end{figure}

\begin{figure}[tbp]
\begin{center}
\vspace{-2cm} \includegraphics[scale=0.6]{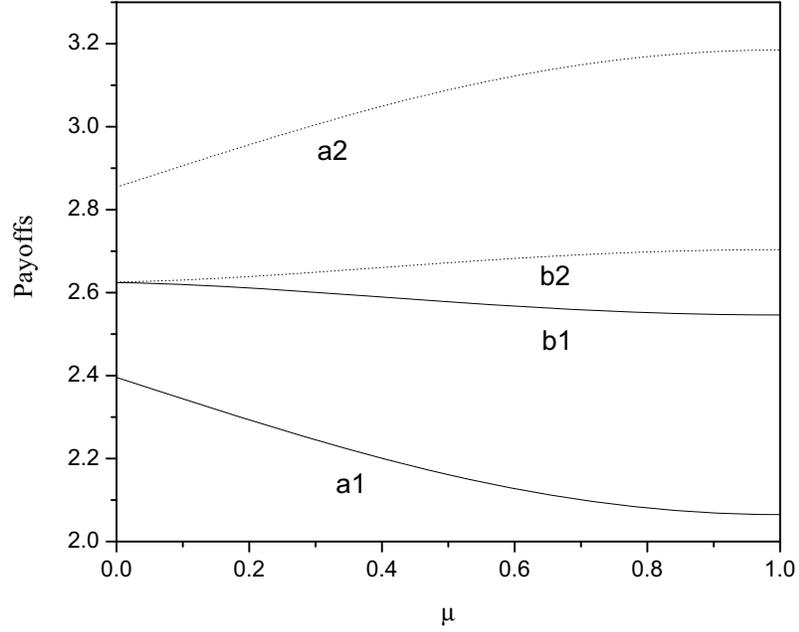} \\[0pt]
\end{center}
\caption{Payoffs of the classical players (Alice /Bob) and the quantum
player (Charlie) are plotted as a function of memory parameter $\protect\mu %
. $ $a_{1}$ and $a_{2}$ are payoffs of the classical players for values of
the decoherence parameter $p=0.7$ and $p=0.3$ respectively. $c_{1}$ and $%
c_{2}$ are payoffs of quantum player for $p=0.7$ and $p=0.3$ respectively.
The other parameters are $\protect\theta _{1}=\protect\theta _{2}=\protect%
\theta _{3}=\protect\pi /2,$ $\protect\beta _{1}=\protect\beta _{2}=\protect%
\alpha _{1}=\protect\alpha _{2}=0,$ $\protect\delta =\protect\gamma =\protect%
\pi /2$ and $\protect\alpha _{3}=\protect\pi /2,$ $\protect\beta _{3}=%
\protect\pi /2$ are the optimal strategies of Charlie.}
\end{figure}

\begin{figure}[tbp]
\begin{center}
\vspace{-2cm} \includegraphics[scale=0.6]{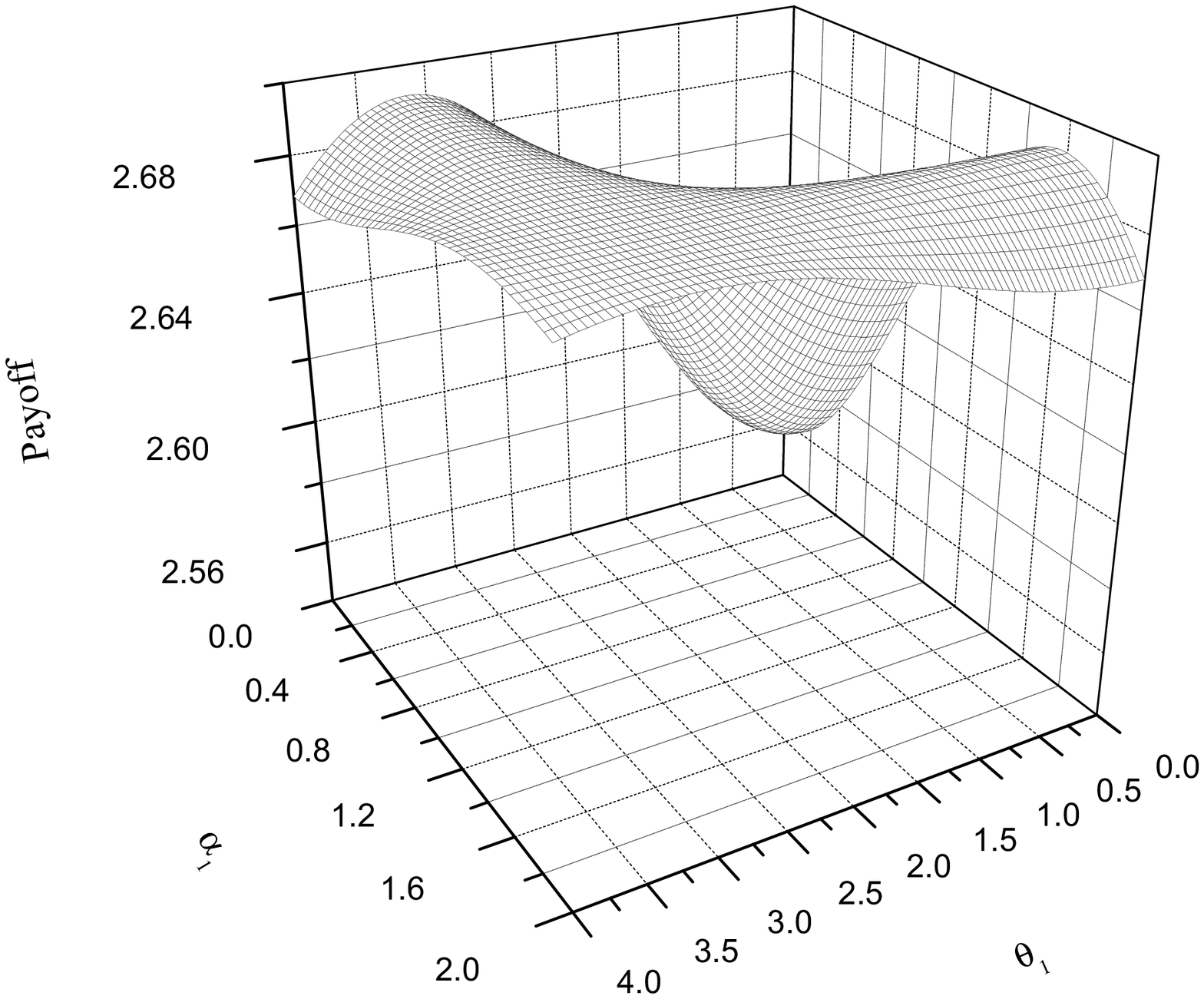} \\[0pt]
\end{center}
\caption{Alice's payoff is plotted as a function of her strategies $\protect%
\alpha _{1}$ and $\protect\theta _{1}$ with $\protect\theta _{2}=\protect%
\theta _{3}=\protect\pi /2,$ $\protect\alpha _{2}=\protect\alpha _{3}=%
\protect\beta _{1}=\protect\beta _{2}=\protect\beta _{3}=0,$ $\protect\delta %
=\protect\gamma =\protect\pi /2$ and $p=\protect\mu =0.3.$}
\end{figure}

\begin{figure}[tbp]
\begin{center}
\vspace{-2cm} \includegraphics[scale=0.6]{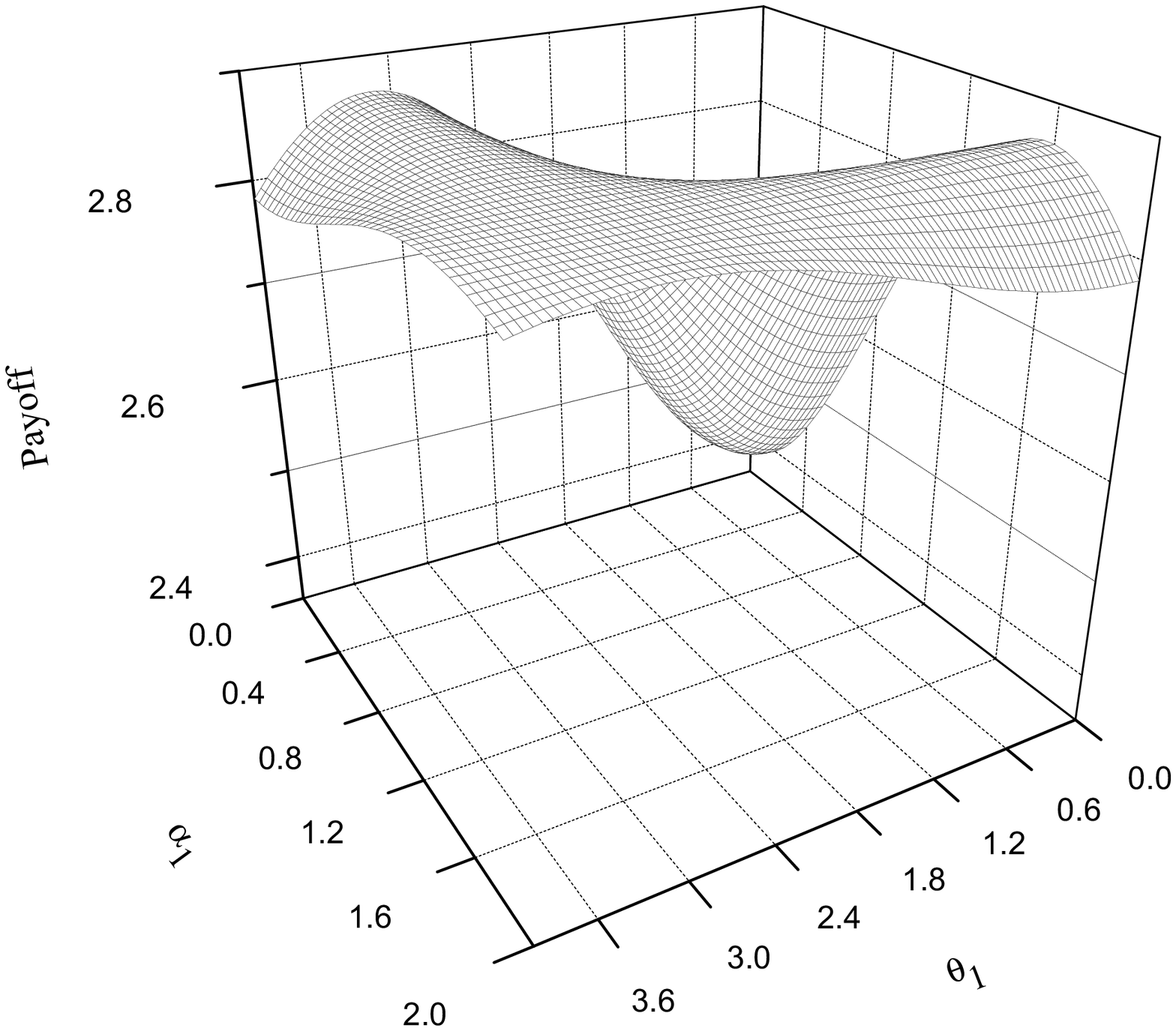} \\[0pt]
\end{center}
\caption{Alice's payoff is plotted as a function of her strategies $\protect%
\alpha _{1}$ and $\protect\theta _{1}$ with $\protect\theta _{2}=\protect%
\theta _{3}=\protect\pi /2,$ $\protect\alpha _{2}=\protect\alpha _{3}=%
\protect\beta _{1}=\protect\beta _{2}=\protect\beta _{3}=0,$ $\protect\delta %
=\protect\gamma =\protect\pi /2$ and $p=\protect\mu =0.7.$}
\end{figure}

\begin{table}[tbh]
\caption{The payoff matrix for the three-player Prisoner's Dilemma game
where the first number in the parenthesis denotes the payoff of Alice, the
second number denotes the payoff of Bob and the third number denotes the
payoff of Charlie.}
\label{di-fit}$%
\begin{tabular}{c}
\hline
\begin{tabular}{ll}
\ \ \ \ \ \ \ \ \ \ \ \ \ \ \ \ \ \ \ \ \ 
\begin{tabular}{l}
\underline{Charlie C} \\ 
\end{tabular}
& \ \ \ \ \ \ \ \ \ \ \ \ \ \ \ \ \ \ \ \ \ 
\begin{tabular}{l}
\underline{Charlie D} \\ 
\end{tabular}
\\ 
\ \ \ \ \ \ \ \ \ \ \ \ \ \ \ \ \ \ \ \ \ \ \ \ 
\begin{tabular}{l}
\\ 
\underline{Bob} \\ 
\end{tabular}
& \ \ \ \ \ \ \ \ \ \ \ \ \ \ \ \ \ \ \ \ \ \ \ \ 
\begin{tabular}{l}
\\ 
\underline{Bob} \\ 
\end{tabular}
\\ 
$%
\begin{tabular}{l}
\\ 
\\ 
\multicolumn{1}{c}{Alice} \\ 
\\ 
\end{tabular}%
\begin{tabular}{l|l}
& \ \ \ C \ \ \ \ \ \ \ \qquad D \\ \hline
\begin{tabular}{l}
C \\ 
\\ 
D%
\end{tabular}
& 
\begin{tabular}{l}
(3,3,3)\qquad (2,5,2) \\ 
\\ 
(5,2,2)\qquad (4,4,0)%
\end{tabular}%
\end{tabular}%
$ & $%
\begin{tabular}{l}
\\ 
\\ 
\multicolumn{1}{c}{Alice} \\ 
\\ 
\end{tabular}%
\begin{tabular}{l|l}
& \ \ \ C \ \ \ \ \ \ \ \qquad D \\ \hline
\begin{tabular}{l}
C \\ 
\\ 
D%
\end{tabular}
& 
\begin{tabular}{l}
(2,2,5)\qquad (0,4,4) \\ 
\\ 
(4,0,4)\qquad (1,1,1)%
\end{tabular}%
\end{tabular}%
$%
\end{tabular}
\\ \hline
\end{tabular}%
$%
\end{table}


\begin{thebibliography}{99}
\bibitem{c} Zurek W\ H 1999 Phys. Today \textbf{44} 36.

\bibitem{bb} Macchiavello C and Palma G M 2002 Phys. Rev. A \textbf{65}
050301.

\bibitem{k} Lidar D A and Whaley K B 2003, Springer Lecturer Notes in
Physics vol \textbf{622} ed Benatti F and Floreanini R (Berlin: Springer) p
83.

\bibitem{Pres} Preskill J 1998 Proc. Roy. Soc. Lond. A \textbf{454\ }385.

\bibitem{Flitney} Flitney A P and Abbott D 2002 Fluct. Noise Lett. \textbf{2}
R175.

\bibitem{Lee} Lee C F and Johnson N 2002 Phys. Rev. A \textbf{67} 022311.

\bibitem{Du} Du, et al. 2003 J. Phys. A: Math. Gen. \textbf{36} 6551.

\bibitem{JOptE} Eisert J and Wilkens M 2000 J. Mod. Opt. \textbf{47} 2543.

\bibitem{DuPLA} Du J, et al. 2001 Phys. Lett. A \textbf{289} 9.

\bibitem{IqbalT} Iqbal A and Toor A\ H 2002 Phys. Lett. A \textbf{300} 537.

\bibitem{e} Meyer D A 1999 Phys. Rev. Lett. \textbf{82} 1052.

\bibitem{f} Eisert J, Wilkins Mand Lewenstein M 1999 Phys. Rev. Lett. 
\textbf{83} 3077.

\bibitem{g} Marinatto L and Weber T 2000 Phys. Lett. A \textbf{272} 291.

\bibitem{Flitney3} Flitney A P and Abbott D 2002 Fluct. Noise Lett. \textbf{2%
} R175.

\bibitem{Pitros} Piotrowski E\ W and Sladkowski J 2002 Physica A \textbf{312}
208.

\bibitem{Pitros2} Piotrowski E\ W and Sladkowski J 2003 Int. J. Theor. Phys. 
\textbf{42} 1089.

\bibitem{zhu} Zhou L, Kuang L\ M 2003 Phys. Lett. A \textbf{315} 426.

\bibitem{a} Du J, Li H, Xu X, Shi M, Wu J, Zhou X, Han R 2002 Phys. Rev.
Lett. \textbf{88} 137902.

\bibitem{b} Prevedel R, Stefanov A, Walther P and. Zeilinger\ A 2007 New J.
Phys. \textbf{9} 205.

\bibitem{I} Chen K L, Ang H, Kiang D, Kwek L C and Lo C F 2003 Phys. Lett. A 
\textbf{316} 317.

\bibitem{m} Nawaz A and Toor A H 2006 J. Phys. A: Math. Gen. \textbf{39}
9321.

\bibitem{n} Ramzan M, Nawaz A, Toor A\ H\ and Khan M K 2008 J. Phys.\ A:
Math. Theor. \textbf{41} 055307.

\bibitem{QMSG} Gawron P and Sladkowski J 2008 Int. J. Quant. Info. \textbf{6}
667.

\bibitem{aaa} Benjamin S\ C and Hayden\ P\ M Phys. Rev. A \textbf{64} (2001)
030301.

\bibitem{II} Du J\ F, et al. 2002 Phys. Lett. A \textbf{302} 229.

\bibitem{Du2} Du J\ F, et al. 2002 Fluct. Noise Lett. \textbf{2} R189.

\bibitem{a3} Flitney A\ P and Abbott D 2004 J. Opt. B: Quantum Semiclass.
Opt. \textbf{6} S860

\bibitem{rr} Cao S, et al. 2007 Chin. Phys. \textbf{16} 915.

\bibitem{h} Kay R, Johnson N F, Benjamin S C 2001 J. Phys. A \textbf{34}
L547.

\bibitem{oo} Yeo Ye and Skeen Andrew 2003 Phys. Rev. A \textbf{67} 064301.

\bibitem{p} Nielsen M A and Chuang I L 2000 Quantum Computation and Quantum
Information (Cambridge: Cambridge University Press).

\bibitem{Flitney2} Flitney A P and Abbot D 2005 J. Phys. A \textbf{38} 449.

\bibitem{qqq} Karimipour V and Memarzadeh L 2006 Phys. Rev. A \textbf{74}
062311.\pagebreak
\end{thebibliography}
\end{document}